\documentclass[journal,article,submit,pdftex,moreauthors]{mdpi} 
\firstpage{1} 
\pubvolume{1}
\issuenum{1}
\articlenumber{0}
\pubyear{2022}
\copyrightyear{2022}
\datereceived{} 
\dateaccepted{} 
\datepublished{} 
\hreflink{https://doi.org/} 
\Title{Jets, Disks and Winds from Spinning Black Holes:\\Nature or Nurture?}
\TitleCitation{Title}

\Author{Roger Blandford $^{1\ast}$\orcidA{} and No\'emie Globus $^{2,3}$\orcidB{}}
\AuthorNames{Roger Blandford and No\'emie Globus}
\AuthorCitation{Blandford, R.; Globus, N.}
\address{$^{1}$ \quad KIPAC, Stanford, CA 94305, USA\\
$^{2}$ \quad Department of Astronomy and Astrophysics, University of California, Santa Cruz, CA 95064, USA\\
$^{3}$ \quad Center for Computational Astrophysics, Flatiron Institute, Simons Foundation, New-York, NY10003, USA}
\corres{Correspondence: rdb3@stanford.edu}
\abstract{A brief summary is given of an alternative interpretation of the Event Horizon Telescope observations of the massive black hole in the nucleus of the nearby galaxy M87. It is proposed that the flow is primarily powered by the black hole rotation, not the release of gravitational energy by the infalling gas. Consequently, the observed millimetre emission is produced by an ``ergomagnetosphere'' that connects the black hole horizon to an ``ejection disk'' from which most of the gas supplied at a remote  ``magnetopause'' is lost through a magnetocentrifugal wind. It is argued that the boundary conditions at high latitude on the magnetopause play a crucial role in the collimation of the relativistic jets. The application of these ideas to other types of source is briefly discussed.}
\keyword{astronomy; galaxies; universe} 
\begin{document}
\section{Introduction}
This essay begins, as does the history of its subject, with the nearby, giant elliptical galaxy M87. The M87 jet was first noticed by Curtis \cite{curtis18}. We now know that such jets are formed by massive black holes residing in the nuclei of distant galaxies \citep{blandford18}. This inference has been validated by remarkable EHT imaging \citep{eht19a}, polarimetry \citep{eht21} and interpreted in the context of a longstanding model for conservative accretion at rates well below critical, where the gas near the black hole forms a torus supported by the pressure of $\sim100\,{\rm MeV}$ ions and observed through synchrotron radiation from $\sim10\,{\rm MeV}$ electrons \citep{eht19b}. This torus confines and collimates two electromagnetic jets which are powered, electromagnetically, by a spinning black hole \citep{blandford77}. 

Although this model may be essentially correct, there are several reasons to consider alternatives. These include:
\begin{itemize}
\item The jet power is $\gtrsim100$ times the observed power from the ring.
\item The mass supply rate at the critical or Bondi radius ($r_B\sim10^6m$, where $m=GM/c^2\sim10^{13}\,{\rm m}$ is the gravitational radius) also seems to exceed greatly that inferred to be accreting on to the black hole.
\item The jets are still being collimated at radii $\sim10^4m$ far beyond the reach of the torus.
\end{itemize}

These concerns suggest to us an alternative model \citep{blandford22} in which a fraction of the power that launches the jets is fed into the accretion disk at small radius and is sufficient to drive off most of the gas supplied around the Bondi radius (or radius of influence) from disk radii, where the binding energy is relatively low compared with that near the black hole \citep{blandford99}. Furthermore, it is proposed that this outflow takes the form a magnetocentrifugal wind \citep{blandford82} launched and collimated by a large scale magnetic field which threads the disk. This wind becomes increasingly toroidal and combines with the dynamical pressure of the gas to collimate the jet \citep{globus16}. The final, essential feature of this model is that gas infalls quasi-spherically. This supplies an important boundary condition and is ultimately responsible for the trapping of magnetic flux and the jet collimation.

\section{Conservation Laws}
Most recent, theoretical research on Active Galactic Nuclei, AGN, has been based upon impressive simulations \citep[e.g.][]{parfrey19}. These have taken a variety of forms - fluid, MHD and kinetic and incorporated radiative transfer, plasma physics as well as general relativity. However, an AGN is much more than the region captured by the EHT image of M87 and it is impossible to achieve the dynamic range needed to span the full range of relevant lengthscales. One approach to connecting simulations across such a large range of scales is to focus on the physics conservation laws and it now seems time to pay more attention to these.

\subsection{Mass}
Black holes are commonly seen to grow by voraciously consuming gas. This is surely the case for the brightest sources, like quasars. However, low power (relative to the Eddington luminosity) sources are generally supposed to accrete mass at a very low rate, much lower than estimates of the supply rate. In the case of M87, the mass supply rate is estimated to be $\sim10^{22}\,{\rm kg\,s}^{-1}$ and the observed mm luminosity is only $\sim10^{34}\,{\rm W}$, very low radiative efficiency for conservative accretion. It is widely supposed that the accretion rate is much smaller than this. If the mass supply at the Bondi radius and accretion rate close to the black hole cannot be reconciled, then there needs to be a strong disk wind to drive away most of the gas and this requires a source of power. 

\subsection{Angular Momentum}
Whatever the rate, the gas supply is generally expected to have sufficient angular momentum within the Bondi radius to form a Keplerian disk. Most of the disk angular momentum resides at large radius and, in a standard, radiative disk, the angular momentum carried inward by the gas, $\propto r^{1/2}$, is balanced by an internal torque $G$ due to magnetic field lying in the disk. If there is a MHD wind, then this can be very efficient in removing angular momentum, as happens with the solar wind. This reduces the magnitude of the internal torque that will be required for a given gas supply.

\subsection{Energy}
Gas flowing inward through a disk carries with it a negative binding energy $\propto r^{-1}$. So, in contrast to the angular momentum, most of the energy resides at small radius. In a standard accretion disk, this energy is liberated and radiated away. However, in addition, the torque that was invoked to balance the flow of angular momentum will do work at a rate $G\Omega$, where $\Omega$ is the Keplerian angular velocity. The divergence of this flow of energy trebles the release of energy and the luminosity. The extra luminosity derives from a reduction of the power radiated at small radius, where most of the energy resides, and requires efficient outward energy transport by the internal torque. If the inflow is adiabatic with small radiative loss, then power is needed to give the gas more than the escape energy and, again, this must be transported outward by the torque in the disk.  Provided that there is still sufficient gas flowing inward at small radius, it can release enough binding energy to account for the extra energy given to the escaping gas.

However, if there are significant radiative losses from the disk  or there is too little mass flow remaining at small radius so that it releases relatively little binding energy, then an additional power source must be invoked beyond the gravitational energy of the accreting gas. (This is reminiscent of the development of solar physics, where nuclear power was invoked to prolong the sun's lifefrom millions  to billions of years.) In this case the natural power source is the spinning black hole. 

Rotational energy can amount to up to $\sim 0.3$ of the black hole mass. In sources like M87, it is invoked to power the jets electromagnetically. In the simplest, stationary, axisymmetric models the power is carried exclusively as a Poynting flux along the jet. However, in less idealized models, only a small fraction of this power suffices to overwhelm the release of binding energy by the inflowing gas and it seems quite likely that essentially all of the mass supplied at large radius is driven off between a few gravitational radii and the radius where an infall transitions to an outflow which we call the magnetopause. This is perhaps a million times larger than the black hole. Where and how this happens depends upon a more detailed physics investigation and the history of the mass supply. Indeed the flow is quite likely to be time-dependent. (It is commonly supposed that the flow is self-similar but this could be a poor approximation.) This is a very different flow of energy from that of an Advection-Dominated Accretion Flow \citep{narayan95} where  essentially all of the mass supply crosses the event horizon but with a large internal energy and, consequently, a low radiative efficiency. Instead, we propose that the gas is always able to cool, the disk is thin and its most important function is to trap and concentrate the magnetic field around the black hole so as to extract its spin energy.

\subsection{Magnetic Flux}
Magnetic field plays a large role in contemporary models of AGN. It takes two basic forms (Fig.~\ref{fig:naturenurture}). Within an accretion disk, it can develop as the nonlinear evolution of the magnetorotational instability. Essentially this involves magnetic turbulence evolving on a dynamical timescale. This produces a tangled magnetic field leading to a non-zero mean value for the Maxwell shear stress tensor component $<T_{r\phi}>$ that is responsible for the torque $G$. The ratio of the shear stress to the total pressure is conventionally called $\alpha$, though simulations typically exhibit more interesting behavior than is captured by a universal, constant value. Similar magnetic field is found in simulations of thick ion tori and is responsible for the synchrotron emission that is observed by EHT. Of course, electrical current accompanies the magnetic field and is computable from its curl, but MHD is most commonly transacted through the magnetic field. As with all forms of turbulence, there will be significant energy dissipated on small scales, especially within a corona. In a hot, collisionless plasma, the form of this dissipation is likely to involve the production of wave modes, especially Alfv\'en waves. It will also involve magnetic reconnection. Both processes are associated with the acceleration of high energy particles, especially relativistic electrons. 

By contrast, the magnetic flux which threads the horizon of the spinning black hole is thought to be slowly varying.  The black hole spacetime acts like a modest electric conductor with an effective resistance $\sim100\,\Omega$. (The associated dissipation occurs, invisibly, behind the event horizon.) This implies that the magnetic flux is not line-tied to the horizon as it is when it threads a spinning neutron star. The field lines end up ``moving'' with an angular velocity about half that associated with the black hole and adjust themselves so as to balance the transverse electromagnetic stress. If we assume that the electromagnetic field adopts the stationarity and axisymmetry of the underlying Kerr spacetime, then there are conserved flows of electrical current, electromagnetic energy and angular momentum along (equipotential and isorotational) magnetic flux surfaces in a non-rotating frame.  Close to the horizon, physical ``observers'' must rotate and if they hover just outside the event horizon, they will see energy flowing into the black hole. 

A useful way to think about the extraction of energy is to note that along a given flux surface rotating with a fixed angular velocity, there will be an inner and an outer light surface. Within the former, a physical observer, on a timelike geodesic must move radially inward; beyond the latter, the motion must be outward. Gravitational energy is not localized within general relativity but, it seems better to consider the electromagnetic power as being extracted from the region between these two surfaces and not from the event horizon.    

To make this more quantitative, a spinning black hole immersed in magnetic flux supported by external toroidal current, and held in place eventually by orbiting gas, generates an EMF which for M87 is roughly $V\sim20\,{\rm EV}$, where E is the SI prefix for $10^{18}$. Given the resistance, an estimate for the associated electrical current per jet is $\sim0.3\,{\rm EA}$. This flow of electromagnetic Poynting flux is what powers the jet and is essentially invisible because the amount of plasma needed to carry the electrical current and produce the electrical charge density is tiny and unlikely to have dynamical consequence. The minimum value of the ratio of the electron pressure needed to support the current to the magnetic pressure, conventionally $\beta$, is of order the ratio of the Larmor radius to the length scale which, in turn is of order the ratio of the rest mass of an electron to $eV$, which is $\sim10^{-13}$. In reality, $\beta$ will be many orders of magnitude greater than this, but this still suggests that that plasma may be dynamically unimportant close to the black hole. In this case, the electromagnetic field may be best described by force-free electrodynamics which essentially equates the divergence of the purely electromagnetic stress-energy tensor to zero. In addition, so little plasma is required to carry electrical current that there may be no need to invoke pair production above the black hole. Instead, sufficient disk plasma may be transported across the magnetic field by small scale magnetic interchange instabilities.

As we move radially outward from the horizon, towards and through the disk, the magnetic field may be either highly ordered, strongly turbulent or somewhere in between. Most descriptions of the accretion disk presume that its large scale magnetic field is constantly regenerated by local disk dynamos. There is no large scale polarity. By contrast, we propose that there is a single polarity maintained over the entire disk, for at least a dynamical time at the infall radius. It can be argued that unipolarity is inevitable. Differential rotational is likely to render the magnetic field approximately axisymmetric. If we move radially inward and the sign of the poloidal magnetic field reverses twice then there will be magnetic reconnection will allow the ring of reversed magnetic flux to be released. Through this mechanism a constanbt sign of high latitude polidal field anchored to the disk can be established from the outside in as most of the magnetic flux threading the disk is found at large disk radius.   This proposal is not inconsistent with there being an active corona, just as happens with the sun. The detailed distribution of the magnetic flux  will reflect  the disk physics and the history. Note that only a tiny fraction of the total magnetic flux threading the disk is needed to power the jets.

It is not known how far out in radius the disk extends. In M87, it has been observed out to $\sim6\times10^4m$ or $\sim20\,{\rm pc}$ where the deprojected orbital speed is $\sim1200\,{\rm km\,s}^{-1}$, consistent with the black hole mass \citep{ford94}. The behavior of the gas beyond this radius is complex and poorly understood despite impressive X-ray imaging by Chandra. One possible scenario is that the inflow from the surrounding cluster is quasi-spherical until its relatively small angular momentum, causes it to hang up and fall into equatorial plane. This is a natural location for a magnetopause --- an interface where the infall meets the wind from the disk in our model. The magnetic state of the wind is hard to guess and we must be guided by observations.

\subsection{Current}
We should also consider the toroidal component of magnetic field around the black hole, in the ergomagnetosphere and in the ejection disc. This is best characterized using the flow of poloidal current, $I$. If the poloidal field and the angular velocity are aligned with the $+z$ direction, then $I$ flows inward from the jet to the horizon at high latitude and, in a steady state, must be part of a stationary circuit. This  current must exit the horizon at lower latitude. (This is achieved by having opposite charges fall in preferentially in the two zones.) The combined current associated with the two jets then flows preferentially along the equatorial plane. Some of this current may complete as a return current flowing the outer boundary of the jets and be associated with Ohmic dissipation/particle acceleration. radio emission --- the observed jet sheath. 

However, there will still be net negative current within cylindrical radius $\varpi$ supporting a toroidal field which allows the disk wind to carry away angular momentum from the disk.If $B_\phi$ falls off slower than $\varpi^{-1}$, then there will be an inward Lorentz force causing the wind to collimate towards the jet axis. Even if $B_\phi\propto\varpi^{-1}$, the radial magnetic stress associated with it will be $(\varpi_{\rm out}/\varpi_{\rm jet})^2$ larger at the jet than at $\varpi_{\rm out}$. In principle, this allows the wind to transmit and amplify stress exerted at the magnetopause onto the jet. Simple pinches are notoriously unstable, but winds are more complicated.  

\section{Ergomagnetosphere}
The basic idea that the spinning black hole not only powers the jets but drives away most of the gas supplied at large radius requires a mechanism for transporting energy radially outward from the black hole across to the disk through the ergomagnetosphere. Several options have been entertained. There may be magnetic flux tubes that connect the inner disk to the more rapidly rotating event horizon. The torque will therefore be positive. As the disk is a much better conductor than the horizon. the field lines should rotate with the angular velocity of the innermost part of the accretion disk. However, they are also likely to be ephemeral as the gas in the disk flows in or away.  

An alternative general mechanism for transmitting the torque supposes that there is vertical magnetic field passing through the equatorial plane between the horizon and the disk. We can also suppose that the angular velocity of these field lines declines steadily with radius. Such a configuration raises an interesting problem for magnetic field lines passing through the ergosphere. In order for the electromagnetic field to be essentially magnetic, more specifically for the Lorentz invariant $B^2-E^2$ to be positive, then it may have to rotate with a finite angular velocity. However this implies that there will be electromagnetic angular momentum and energy carried away along these field lines with no obvious source. We have conjectured that what actually happens is that the magnetic field lines will indeed decelerate to some minimum angular velocity, where $E\rightarrow B$ and that there is a transition to a form of electromagnetic turbulence where random force-free waves are created on negative energy orbits and these propagate backwards into the horizon, thereby maintaining conservation of angular momentum and energy \citep{penrose69}. Through this device, even more power can be extracted from a spinning black hole than by the magnetic flux that threads the horizon directly.

However, this mechanism does not transmit power outward to the disk. Here, an additional process must be invoked, somewhat similar to the magnetorotational instability. It appears that differentially rotating magnetic flux surfaces passing through the ergomagnetosphere are subject to non-axisymmetric, force-free instability and the nonlinear development of this instability can quite plausibly divert a few percent of the jet power to drive a powerful disk wind as required. It should be noted that a MHD disk wind passes through an Alfv\'en critical surface, typically at some height above the disk. The flow below this surface is causally connected to the disk and so any angular momentum and energy transmitted to it --- a much larger target --- is effectively communicated to the disk.  

\begin{figure}[H]
\begin{adjustwidth}{-\extralength}{0cm}
\centering
\includegraphics[width=15.5cm]{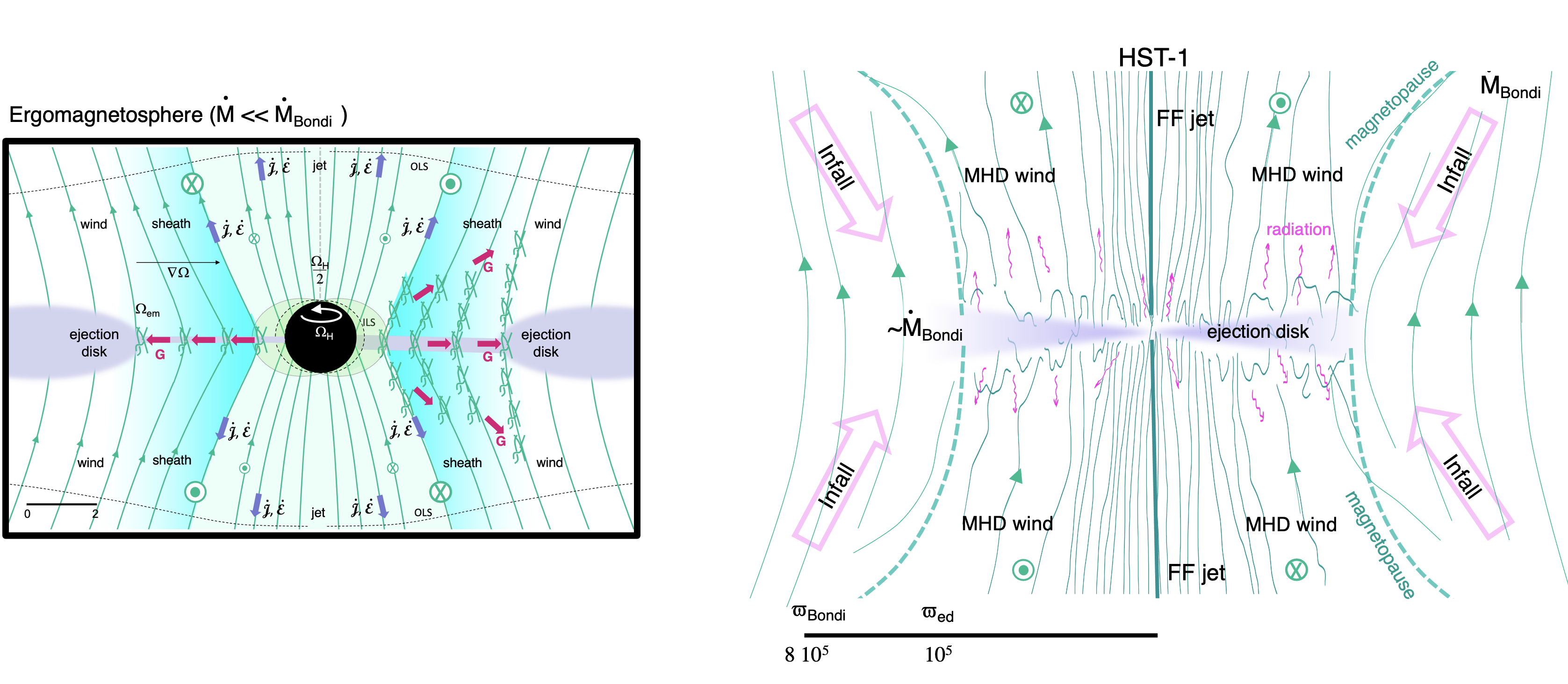}
\includegraphics[width=15.5cm]{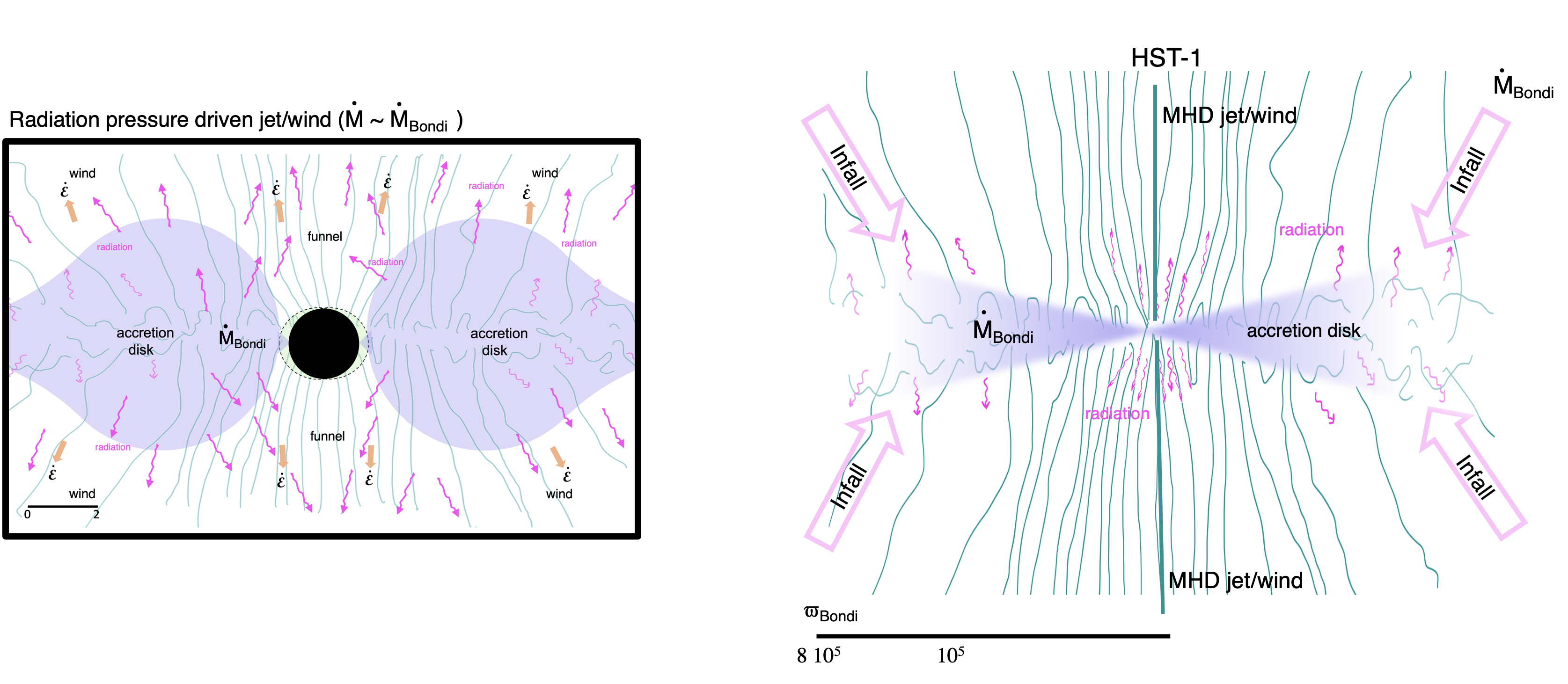}
\end{adjustwidth}
\caption{{\bf Upper Left.} Schematic illustration of a model of jet formation in M87. The jet is powered by a rapidly spinning, massive black hole (nature) in the nucleus of a giant elliptical galaxy located in the Virgo cluster. The rotational energy is primarily extracted by magnetic flux which threads the event horizon and which lies in the middle of a larger, unsteady ergomagnetosphere which, in turn, is retained by the inertia of orbiting gas in a thin ejection disk. A smaller power is also transmitted laterally from the black hole to the ejection disk through the ergomagnetosphere. {\bf Upper Right.} Considered on the larger scale, the power that is applied to the inner boundary of the ejection disk, is transmitted outward through the action of a magnetic torque $G$ acting within the disk. This transports energy and angular momentum outwards at a high enough rate to unbound most of the gas that is supplied to the disk at its outer radius, $\sim10^{5-6}$ black hole gravitational radii. The escape of the gas from the disk is mediated by large-scale vertical magnetic field with a single polarity extending over all radii. Gas that is attached this vertical magnetic field will be flung outward by centrifugal force and its inertia will, in turn, drag the field lines backward and create a toroidal component to the field, which collimates the outflow towards the axis of rotation. This MHD wind is ultimately held in place by the ram pressure of the infalling gas at high latitude. The boundary between the wind and the infall is known as the magnetopause. {\bf Lower Left.} A more conventional model of jet formation in M87 invokes an approximately conservative mass flow from infall to the black hole. The jet could still be powered by the black hole spin or, if the hole is slowly rotating, driven by gas pressure and collimated by a thick torus of orbiting gas supported primarily by hot ions.  The emission is powered by the release of gravitational binding energy (nurture). {\bf Lower Right.} The torus is orbited by a thinner accretion disk from which the luminosity also derives from the gravitational binding energy of the infalling gas and exceeds any wind which may also be driven the thermal pressure and radiation. A similar description pertains to models of super-Eddington accretion where radiation pressure replaces gas pressure (Sec.~4).}
\label{fig:naturenurture}
\end{figure}  
\section{Other Sources}
So far we have focused on the particular and important case of M87. However, the ideas outlined here could have relevance for other astronomical sources. The most basic question is what is needed to form a jet. We have suggested that, in M87, there is a central source of power, a unidirectional field threading a disk and confinement for high latitude at the magnetopause separating the outflow from the infall. However, these features may not all be necessary and many variations are possible \citep{konigl94}. 

A second source studied with comparable resolution measured in gravitational radii by EHT is Sgr A$^*$ \citep{eht22}. A ring of gas is observed which, it is argued, has an angular velocity pointed, (surprisingly), towards us and perhaps three brightness peaks, although the nature and reality of these peaks seems uncertain.  Sgr A$^*$ is an extremely dim source, with luminosity $\sim10^{29}\,{\rm W}$, mostly in the sub mm band. With a "conventional" radiative efficiency of $\sim0.1c^2$, a mass accretion rate of only $\sim10^{13}\,{\rm kg\,s}^{-1}$ would suffice. However, mass supply rates as large as $10^{18}\,{\rm kg\,s}^{-1}$ have been proposed, either from stellar mass loss or from an orbiting ring of molecular gas. Even if the accretion is quite radiatively inefficient, as commonly supposed, it seems hard to avoid the conclusion that most of the mass supplied is not accreted by the black hole and is instead driven off. One possibility is that the actual accretion rate is $\sim 10^{15}\,{\rm kg\,s}^{-1}$ with radiative efficiency $\sim10^{-3}c^2$ and that a dynamical power released by the binding energy of the gas that does accrete is communicated outward through the disk and drives as much as $10^{18}\,{\rm kg\,s}^{-1}$ from larger radii where the binding energy is lower. The black hole is essentially a passive source of gravitation. Alternatively if could be spinning and magnetized and power the outflow, much as we have suggested is happening in M87. There is no sign of a conventional jet and this could reflect the angular distribution of the gas supply.

Let us consider other types of AGN. When the mass supply is within a range spanning $\sim10^{2-3}$ centered on the Eddington rate, it is conventionally assumed that these disks are able to inflow under magnetic viscosity and radiate efficiently. However, When the mass supply rate is very large, the accreting gas will become radiation-dominated at small radii and a torus supported by radiation pressure \citep{abramowicz13} will form similar to the ion torus (Fig.~\ref{fig:naturenurture}). The trapped radiation cannot escape the plasma, which has two choices. Either it can fall into the black hole or it can escape in a radiation-driven wind. If the central black hole is spinning and magnetized, then outflow is more likely.  Disk winds are likely to be radiation-driven by lines or continuum emission, but may also be magnetized. Indeed, magnetic field may play an important role in propelling and confining the emission line clouds that allow quasars and Seyfert galaxies to be identified \citep{emmering92}. Most of these sources are radio-quiet, generally without prominent jets. Perhaps the gas inflow onto these objects lies mainly in the equatorial plane and the boundary conditions at large radius do not promote magnetic trapping and jet collimation. By contrast, radio-loud quasars with their relativistic jets, may have quasi-spherical gas inflow. JWST observations of many of these sources should be highly instructive.

Galactic superluminal sources \citep{mirabel99} have binary companions which can supply gas either through a Lagrange point or a wind. The latter case may be conducive to jet formation. A more extreme case is provided by Gamma Ray Bursts where the newly formed jet punches a way out through a collapsing stellar envelope or the debris left over from neutron star merger, both circumstances conducive to jet formation \citep{meszaros19}. Tidal Disruption Events of stars by massive black holes can also create jets \citep{dai21} and similar considerations may apply to the fate of the iinfalling gas.  

A much different example is provided by protostars, which have disks and some of these create impressive winds and jets \citep{bally16}. The magnetocentrifugal mechanism is commonly invoked to drive a disk wind, but the central star typically has a sub-Keplerian angular velocity. If the disk extends down to the star, then a boundary layer will be formed with a positive angular velocity gradient. The kinetic energy released in this deceleration is inadequate to drive away all of the gas flowing through the disk. However, if the star is magnetized and the Alfv\'en radius, where the momentum flux in the gas balances the magnetic stress, is larger than the corotation radius, then it is possible for the star to exert a significant, positive torque on the disk. However, the associated power will be larger than in the black hole case because the relevant angular velocity will be that of the star not the disk. It is possible that accretion onto the star will terminate when this happens. Both the star and the disk might dominate the jet power, a choice that is reminiscent of the FRI/II dichotomy. The infall of gas from the surrounding molecular cloud is likely to extend to high latitude and may be crucial for jet collimation. 

Young pulsar wind nebulae, for example those associated with the Crab and Vela pulsars, also form jets \citep{porth17}. These two have high latitude external stress imposed by the expanding supernova remnant at the contact discontinuity separating the shocked pulsar wind from the supernova ejecta. The magnetic field in the Crab Nebula does appear to be highly organized from X-ray polarimetry and this can lead to jet collimation. Of course, there is no disk but there is an equatorial energy outflow associated with the low latitude relativistic outflow from the pulsar.

\section{Discussion}
In this brief essay, we have summarized a heterodox interpretation of jet formation inspired by the EHT observations of M87. We have argued that the dominant power in this source and perhaps several other source classes, derives from a central spinning, magnetized black hole rather than the gravitational energy released by the accreting gas --- nature not nurture. it is clear that observational capabilities are developing fast and we can look forward to developing a much more complete picture of accretion onto black holes under a wider range of circumstances.
\vspace{6pt} 

\funding{NG’s research is supported by the Simons Foundation, the Chancellor Fellowship at UCSC and the Vera Rubin Presidential Chair.}
\acknowledgments{We thank our many collaborators, past and present for their ideas, encouragement and critique over many years.}
\conflictsofinterest{The authors declare no conflict of interest.} 
\begin{adjustwidth}{-\extralength}{0cm}
\reftitle{References}

\end{adjustwidth}
\end{document}